\documentclass[12pt,aip,jcp]{revtex4-1}
\usepackage{amsmath}
\usepackage{amssymb}
\usepackage[dvips]{graphicx}
\usepackage[svgnames]{xcolor}
\usepackage[colorlinks=true,citecolor=blue,linkcolor=blue,citebordercolor=white,linkbordercolor=white]{hyperref}
\usepackage{natbib}

\def\eq#1{{Eq. (\ref{#1})}}

\begin{document}

\title[Hopping charge transport in amorphous semiconductors]{Hopping charge transport in amorphous semiconductors with the spatially correlated exponential density of states}

\author{S.V. Novikov}
\email{novikov@elchem.ac.ru}
\affiliation{A.N. Frumkin Institute of
Physical Chemistry and Electrochemistry, Leninsky prosp. 31,
119071 Moscow, Russia}
\affiliation{National Research University Higher School of Economics, Myasnitskaya Ulitsa 20, Moscow 101000, Russia}



\begin{abstract}
Hopping charge transport in amorphous semiconductors having spatially correlated exponential density of states has been considered. Average carrier velocity is exactly calculated for the quasi-equilibrium (nondispersive) transport regime. We suggest also a heuristic approach for the consideration of the carrier velocity for the non-equilibrium dispersive regime. \end{abstract}

\maketitle

\section{Introduction}

Carrier hopping between localized states is one of the most important charge transport mechanisms in amorphous semiconductors. Here we consider highly disordered semiconductors having disorder high enough to provide a full localization of all relevant states. There is a general agreement that this picture fairly well describes typical organic amorphous semiconductors due to rather weak intermolecular interaction. In fact, intermolecular interactions are frequently so weak and resulting bands in organic crystals are so narrow that at the room temperature the band transport is destroyed by lattice vibrations and the dominant transport mechanism is hopping even in crystals.\cite{Gershenson:973,Ortmann:023011}

Hopping microscopic dynamics is dictated by the hopping rate which depends on energy difference $\Delta U_{ij}=U_j-U_i$ and distance  $\vec{r}_{ij}=\vec{r}_j-\vec{r}_i$ between the initial $i$ and final $j$ transport sites, $\Gamma=\Gamma(\Delta U_{ij},\vec{r}_{ij})$. Electric field $\vec{E}$ affects charge transport mostly by shifting the energy difference $\Delta U_{ij}\Rightarrow \Delta U_{ij}-e\vec{E}\vec{r}_{ij}$, which inevitably requires a spatial displacement of the carrier along the field direction. This means that the field dependence of the drift mobility $\mu$ is determined by the interplay of the probability density for $U$, i.e., the density of states (DOS) $g(U)$, and spatial correlation of the random energy landscape $U(\vec{r})$ which dictates a typical energy difference for a given distance $r$.

Effect of spatial correlation on the charge transport is fairly well studied for the case of Gaussian DOS. There is a common belief that such DOS is typical for many amorphous organic semiconductors containing high concentration of molecules having permanent dipole or quadrupole moments.  \cite{Bassler:15,Dieckmann:8136,Novikov:14573,Dunlap:542,Novikov:2584} Long range interaction of charge carriers with randomly located dipoles and quadrupoles easily produces a Gaussian DOS because the contribution from many sources naturally implies applicability of the central limit theorem. In addition, such sources inevitably produce long range spatial correlation of the energy landscape $U(\vec{r})$. \cite{Novikov:14573,Dunlap:542}
Direct observation of the cluster structure of the correlated energy landscape is beyond possibilities of the modern microscopic technique, yet the particular type of the energy correlation function $C(\vec{r})=\left<U(\vec{r})U(0)\right>\propto 1/r$, expected in polar organic materials, provides the most natural explanation of the so-called  Poole-Frenkel mobility field dependence $\ln\mu\propto E^{1/2}$, ubiquitous in amorphous organic materials. \cite{Dunlap:542,Novikov:1167}

At the same time, some authors argue that there is an evidence for the exponential tail of the DOS
\begin{equation}\label{exp-DOS}
g(U)=\frac{N_0}{U_0}e^{U/U_0},\hskip10pt U < 0,
\end{equation}
in some amorphous organic materials, here $N_0$ is the total density of transport sites.  \cite{Vissenberg:12964,Tachiya:085201,Street:165207,Schubert:024203}
Development of such DOS was also observed in the computer simulation of the amorphous conjugated polymer  poly-3-hexylthiophene.\cite{Frost:255} Exponential DOS is usually associated with amorphous inorganic semiconductors,\cite{Mott:book} where it is formed in the band gap as a localized tail to the band of delocalized states. We will discuss briefly in Section \ref{sec-exp} to what extent our results could be applied to the materials having delocalized states in addition to the exponential tail of localized states.

For the exponential DOS transport properties are very different from the case of the Gaussian DOS. The most striking difference is that for low temperature $kT/U_0 < 1$ carriers do not attain a quasi-equilibrium state with constant average velocity but the carrier velocity monotonously decreases with time and, hence, with the thickness $L$ of the transport layer as
\begin{equation}\label{disp-L}
v_L\propto 1/L^{\frac{1}{\alpha}-1}, \hskip10pt \alpha=kT/U_0,
\end{equation}
while for $kT/U_0 > 1$ a quasi-equilibrium regime with constant velocity eventually develops (for the Gaussian DOS the  quasi-equilibrium regime develops for $t\rightarrow\infty$ at any temperature). \cite{Rudenko:177,Arkhipov:189}

Transport properties for the spatially correlated exponential DOS have not been studied yet. In this paper we consider effect of spatial correlation of the random energy landscape for the case of the exponential DOS. Long time features of the hopping transport are well described by the approximation based on the diffusion equation for the carriers moving in the random energy landscape $U(\vec{r})$. We consider the 1D case, where behavior of the carrier density $n(x,t)$ is governed by the equation
\begin{equation}\label{diffusion}
\frac{\partial n}{\partial t}=D_0 \frac{\partial}{\partial x}\left[\frac{\partial n}{\partial x}+\frac{1}{kT}\left(\frac{\partial U}
{\partial x}-eE\right)n\right].
\end{equation}
Here $D_0$ is a bare diffusion coefficient in the absence of disorder. We consider the case where carrier density is very low and the effect of filling of the transport sites can be neglected.  For the periodic boundary conditions the carrier velocity for the stationary case is
\begin{equation}\label{stat-v}
v_L=\frac{D_0\left(1-e^{-\gamma L}\right)}{\int\limits_0^L dx \hskip2pt\exp\left(-\gamma x\right)Z(x,L)}, \hskip10pt Z(x,L)=\frac{1}{L}\int\limits_0^L dy \hskip2pt\exp\left[\frac{U(y)-U(x+y)}{kT}\right], \hskip10pt \gamma=v_0/D_0,
\end{equation}
here $v_0=eE D_0/kT$ is a bare carrier velocity in the absence of disorder (in this case the Einstein relation is certainly valid). \cite{Parris:2803,Parris:5295} In this paper we are going to provide an exact solution for the case of the nondispersive quasi-equilibrium transport where all transport parameters become constants for $L\rightarrow\infty$. In particular, for the infinite medium the average carrier velocity goes to
\begin{equation}\label{stat-v_inf}
v=\frac{D_0}{\int\limits_0^\infty dx \hskip2pt\exp\left(-\gamma x\right)Z(x)}, \hskip10pt Z(x)=\lim\limits_{L\rightarrow\infty}Z(x,L)=\left<\hskip2pt\exp\left[\frac{U(0)-U(x)}{kT}\right]\right>,
\end{equation}
where angular brackets mean statistical averaging over realization of $U(x)$ which is effectively provided by the consideration of the infinite sample. We present an effective way to calculate the statistical average $Z(x)$ for the nondispersive case $U_0/kT < 1$ and then calculate the mobility field dependence for various cases of the correlated exponential distribution of $U$. We suggest also a heuristic approach to estimate this dependence for the dispersive non-equilibrium transport regime $U_0/kT > 1$.

Usually the shape of the DOS is not exactly exponential, having only the exponential tail for low energies.  We may assume that there is some additional density at $U>0$ which is not described by \eq{exp-DOS}. We will see that the most interesting results for the carrier dynamics in the nondispersive regime could be obtained for $kT\rightarrow U_0$. In this case the details of the DOS for high energies are not very important. Indeed, in the quasi-equilibrium regime the average carrier energy for the exponential DOS is
\begin{equation}\label{Uavg}
\left<U\right>=\frac{1}{U_T}\int\limits_{-\infty}^0 dU \hskip2pt U\exp\left(\frac{U}{U_0}-\frac{U}{kT}\right)=-U_T\rightarrow -\infty,\hskip10pt U_T=\frac{U_0 kT}{kT-U_0},
\end{equation}
and the fraction of carriers $p\left(U_c\right)$ residing below some energy $U_c < 0$
\begin{equation}\label{fraction}
p\left(U_c\right)=\frac{1}{U_T}\int\limits_{-\infty}^{U_c} dU \hskip2pt \exp\left(\frac{U}{U_T}\right)=\exp\left(U_c/U_T\right) \end{equation}
goes to 1 for any $U_c$ if $kT$ is sufficiently close to $U_0$. Analogously, contribution of states with $U > 0$ is not important for $Z(x)$. Deviation of the localized DOS from the pure exponential form for higher energy gives insignificant correction for any reasonable shape of the DOS at $U>0$ and, hence, is not important for our results at the vicinity of $T=U_0/k$. Contribution from the states with $U>0$ is even less important for the dispersive transport regime $kT<U_0$.

\section{Gaussian representation for a spatially correlated exponential random energy landscape}

To calculate $Z(x)$ for a wide class of the spatially correlated exponential distributions we are going to use a trick well known in the probability theory. Indeed, let
\begin{equation}
U=-\frac{1}{2}U_0\left(X^2+Y^2\right),
\label{U-XY_a}
\end{equation}
where $X$ and $Y$ are two independent identically distributed Gaussian random variables with zero mean and unit variance $\sigma^2=1$. Then the distribution of $U$ is described by \eq{exp-DOS}.  \cite{Devroye-book} If $X$ and $Y$ are Gaussian variables with correlation coefficients $c_X$ and $c_Y$, then the bivariate distribution for $X$ has a form
\begin{equation}\label{2a-PDF_GX}
P_G(X_1,X_2,c_X)=\frac{1}{2\pi\sqrt{1-c^2_X}}\exp\left[
-\frac{X_1^2+X_2^2-2c_X X_1X_2}{2\left(1-c^2_X\right)}\right],
\end{equation}
(and the similar distribution for $Y$). \cite{Feller:book}
Quadratic form in the exponent of \eq{2a-PDF_GX} is positively defined if $|c_X|<1$, but if we consider the correlation coefficient $c_X$ as a correlation function of some physical field $X(x)$, i.e., $c_X=c_X(x)$, then a more natural  assumption is $0<c_X<1$. For $x=0$ we have $c_{X,Y}(0)=1$ and for $x\rightarrow\infty$ $c_{X,Y}(x)\rightarrow 0$. Using this trick the averaging over $U(x)$ is replaced by the averaging over $X(x)$ and $Y(x)$.

The binary correlation function for the exponential random field is
\begin{equation}\label{2a-corrU}
c_U(x)=\left<U(x_1) U(x_2)\right>-\left<U\right>^2=\frac{U_0^2}{2}\left[c_X^2(x)+c_Y^2(x)\right],\hskip10pt x=x_1-x_2,
\end{equation}
so in this way we can model any positive binary correlation function for the random field $U(x)$. At the same time, it is important to note that using this approach it is not possible to generate an arbitrary feasible correlated exponential random distribution. Indeed, for the Gaussian random field all correlation functions can be expressed by the binary correlation function. This is a unique feature of the Gaussian random field, it does not hold for other random fields, such as the exponential field. Obviously, in our ansatz all correlation functions of the resulting field $U(x)$ can be expressed using the correlation functions $c_X$ and $c_Y$ of the initial Gaussian random fields $X(x)$ and $Y(x)$. This is not true for an arbitrary exponential random field. Hence, our approach covers only a limited subset of all exponential random fields. At the same time, our approach may have a physical justification in some amorphous organic materials; indeed, contribution from the molecular polarizability describes the random energy as a sum of squares of three Gaussian variables giving the density of states $g(U)\propto \left(-U\right)^{1/2}\exp\left(U/U_0\right)$, very close to the exponential one. \cite{May:136401} Polarizability mechanism gives $c_U(x)\propto 1/x^5$ and the possibility to provide a more long range correlation is an open question.

A natural generalization of the representation (\ref{2a-PDF_GX}) could be
\begin{equation}\label{2a-gen-PDF_G}
P(X_1,X_2)=\int\limits_{-1}^1 dc W(c)P_G(X_1,X_2,c), \hskip10pt \int\limits_{-1}^1 dc W(c)=1,
\end{equation}
(and the similar representation for $P(Y_1,Y_2)$). Again, in most reasonable cases the weight function $W(c)=0$ for $c<0$. It is easy to check that
\begin{equation}\label{2a-int-gen-PDF_G}
\int\limits_{-\infty}^\infty dX_2 P(X_1,X_2)=\frac{1}{\sqrt{2\pi}}\exp\left(-\frac{X_1^2}{2}\right),
\end{equation}
and the analogous relation is true for $\int\limits_{-\infty}^\infty dX_1 P(X_1,X_2)$. Hence, $P(X_1,X_2)$, though being of the non-Gaussian form,  provides the proper one-point Gaussian distributions for $X_1$ and $X_2$ and, as a result, the proper exponential distribution for $U$. For the non-negativity of $P(X_1,X_2)$ it is sufficient to demand $W(c)\ge 0$ everywhere. Probably, this is the necessary condition, too. At least, for the simplest test case
\begin{equation}\label{2a-2c}
W(c)=A\delta(c-c_1)+(1-A)\delta(c-c_2)
\end{equation}
the corresponding function $P(X_1,X_2)$ is positive everywhere only for $0\le A\le 1$. In this paper we are going to limit our consideration to the simplest case described by \eq{2a-PDF_GX} with $c_X=c_Y=c(x)$.

A direct calculation gives for $Z(x)$
\begin{equation}
Z(x)=\frac{1}{1-\varkappa^2\left[1-c^2(x)\right]}, \hskip10pt \varkappa=U_0/kT
\label{Z}
\end{equation}
for $c_X=c_Y=c(x)$, and a more general relation for the representation (\ref{2a-gen-PDF_G})
\begin{equation}\label{<exp>-g}
Z(x)=\int\limits_{-1}^1\frac{dc_X W(c_X)}{\left[1-\varkappa^2\left(1-c^2_X\right)\right]^{1/2}}
\int\limits_{-1}^1\frac{dc_Y W(c_Y)}{\left[1-\varkappa^2\left(1-c^2_Y\right)\right]^{1/2}}.
\end{equation}

Essentially, we may consider the use of the auxiliary variables $X(x)$ and $Y(x)$ as a technical trick to derive a relation between the correlation function $Z(x)$ and  correlation characteristics of the exponential random energy landscape $U(x)$. It turns out that in the simplest case of \eq{Z} the only relevant characteristic of $U(x)$ is its binary correlation function $c_U(x)=U_0^2c^2(x)$, while in other situations the connection between $Z(x)$ and $U(x)$ becomes more intricate. Yet, as it was mentioned earlier in connection with Ref. \onlinecite{May:136401}, in some situations variables $X(x)$ and $Y(x)$ (or analogous ones) could be realized as true physical random fields providing the necessary exponential or near-exponential distribution of $U(x)$.

\section{Mobility field dependence: nondispersive regime}
\label{ND}

Let us consider general features of the mobility field dependence for the exponential DOS. Here and later while considering the mobility field dependence we are always going to use the dependence of the dimensionless ratio $v/v_0=\mu/\mu_0$ on $v_0=\mu_0 E$ instead of $\mu(E)$, here $\mu_0$ is a carrier mobility in the absence of disorder. For $v_0\rightarrow 0$ (i.e., for $E\rightarrow 0$) the integral in \eq{stat-v_inf} is dominated by the large distance $x\rightarrow\infty$, where $c(x)\rightarrow 0$ and
\begin{equation}\label{2a-Zinf2}
Z(\infty)=\frac{1}{1-\varkappa^2},
\end{equation}
so
\begin{equation}\label{2a-v1_E->0}
v(E\rightarrow 0)\approx v_0\left(1-\varkappa^2\right).
\end{equation}
If $\varkappa\rightarrow 1$, then $Z(x)$ diverges and carrier velocity for the infinite layer goes to $0$ signalling the transition to the dispersive transport regime where the average velocity monotonously decays with the increase of $L$. Here we see the exact agreement with the multiple trapping (MT) model of the carrier transport. \cite{Arkhipov:189} In the opposite limit $v_0 \rightarrow \infty$ (or $E\rightarrow \infty$) the integral (\ref{stat-v_inf}) is dominated by $x\rightarrow 0$ where $Z(x)\approx 1$ and
\begin{equation}\label{2a-v1_E->infty}
v(E\rightarrow \infty)\approx v_0.
\end{equation}
Comparison of \eq{2a-v1_E->0} with \eq{2a-v1_E->infty} immediately gives us a very important conclusion: in order to make a reliable comparison of the experimental data for the nondispersive regime with the theoretical formula one have to carry out experiments for the temperature very close to the transition to the dispersive regime, $kT/U_0\rightarrow 1$. Indeed, a proper functional form of the mobility field dependence could be extracted from experimental data only when the variation of the mobility over the tested field range is significant. Such variation is restricted from above by the ratio $v(E\rightarrow \infty)/v(E\rightarrow 0)=1/(1-\varkappa^2)$. Hence, $1-\varkappa^2$ should be very close to 0. A difficulty to maintain the prescribed temperature with high accuracy is a major obstacle for experimental study of correlation effects in the nondispersive regime.

Let us consider a possible functional form of the mobility field dependence for different types of the correlation function $c^2(x)$. The simplest yet important case is the short range correlated disorder with $c^2(x)=\theta(a-x)$, where $\theta(x)$ is a unit step function. In this case
\begin{equation}
v=v_0 \left(1+\frac{\varkappa^2}{1-\varkappa^2}e^{-\gamma a}\right)^{-1}.
\label{2a-Short-Range}
\end{equation}
We should note that the relation $v\propto\exp(\gamma a)$ for the moderate field is the intrinsic feature of the short range spatial correlation, it is observed also for the Gaussian DOS. \cite{Novikov:2532}

\begin{figure}[tbp]
\includegraphics[width=3in]{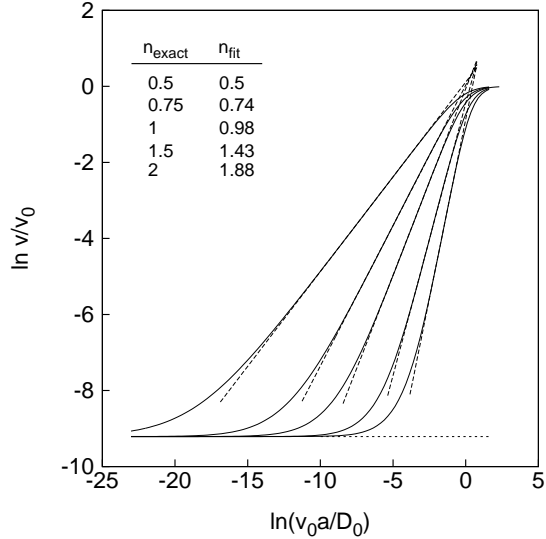}
\caption{Mobility field dependence (solid lines) calculated using \eq{stat-v_inf} for the power law correlation function $c^2(x)=a^n/(x^2+a^2)^{n/2}$ for various values of $n$: 0.5, 0.75, 1, 1.5, and 2, from the upmost curve to the bottom, correspondingly, and for $\delta=1-\varkappa^2=1\times 10^{-4}$. Horizontal dotted line shows the limiting dependence $v/v_0=1-\varkappa^2$, and the broken lines show fits for power law dependence $v/v_0\propto v_0^{n}$.} \label{XY_fig1}
\end{figure}

\begin{figure}[tbp]
\includegraphics[width=3in]{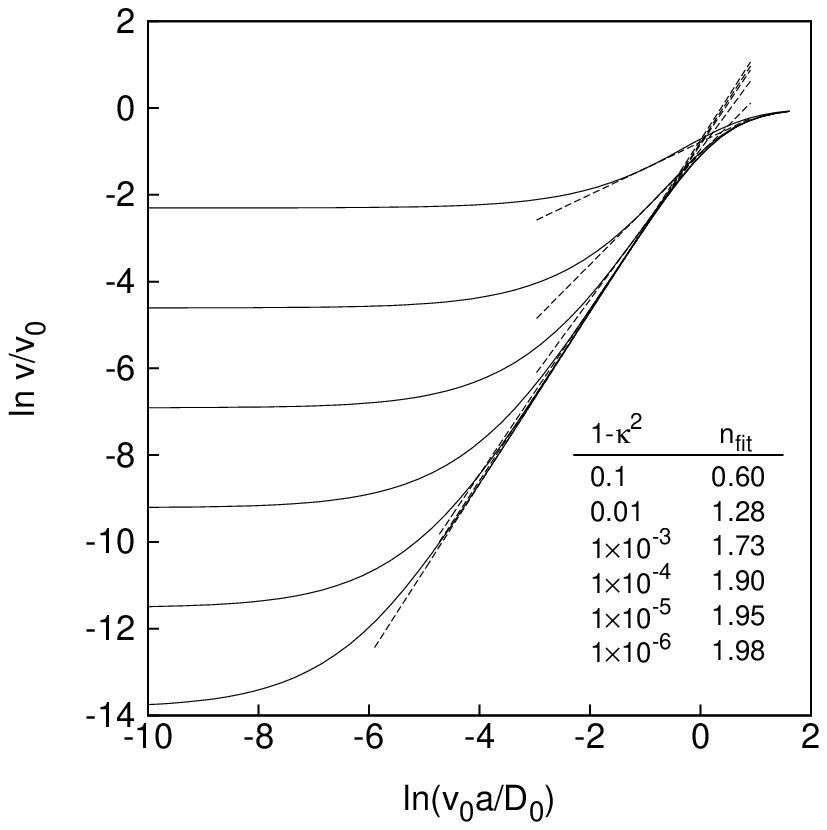}
\caption{Mobility field dependence (solid lines) calculated using \eq{stat-v_inf} for the power law correlation function $c^2(x)=a^2/(x^2+a^2)$ (i.e., $n=2$) for various values of $1-\varkappa^2$: 0.1, 0.01, $1\times 10^{-3}$, $1\times 10^{-4}$, $1\times 10^{-5}$, and $1\times 10^{-6}$, from the upmost curve to the bottom, correspondingly. Effective values of $n_{\rm fit}$ for different $\varkappa^2$ are estimated from the best fit of the middle linear regions of the curves to the relation $v/v_0 \propto (\gamma a)^n$ (broken lines). } \label{XY_fig2}
\end{figure}

Another important case is the power law correlation function $c^2(x)=a^n/(x^2+a^2)^{n/2}$. Such correlation functions are typical for organic materials. \cite{Novikov:2584} In the close vicinity of the transition to the dispersive regime $1-\varkappa^2=\delta \ll 1$ the integral in \eq{stat-v_inf} could be easily estimated
\begin{equation}
\int\limits_0^\infty dx e^{-\gamma x}Z(x)\approx
\frac{1}{a^n}\int\limits_0^\infty dx e^{-\gamma x}x^n = \frac{\Gamma(n+1)}{\gamma (a\gamma)^n}.
\label{2a-Sp}
\end{equation}
This estimation is valid if $\delta \ll c^2(x_m)$ and $x_m \gg a$, where $x_m$ is the position of the maximum of the function $\exp(-\gamma x)\left(x^2+a^2\right)^{n/2}$, leading to the final condition $n\delta^{1/n}\ll \gamma a\ll n$. Average carrier velocity is
\begin{equation}
v\simeq \frac{v_0}{\Gamma(n+1)}\left(\gamma a \right)^n
\label{2a-Sp3}
\end{equation}
in good agreement with the numerical calculation of the integral (\ref{stat-v_inf}) (see Fig. \ref{XY_fig1}). Coefficient of proportionality in the relation $v\simeq A v_0 (\gamma a)^n$, calculated for the data presented in Fig. \ref{XY_fig1}, agrees with \eq{2a-Sp3} with the reasonable accuracy around 10\%. Fig. \ref{XY_fig2} illustrates the necessity to measure experimental mobility field dependence in a very close vicinity of the critical point $\varkappa=1$ in order to extract the reliable value of $n$.

\begin{figure}[tbp]
\includegraphics[width=3in]{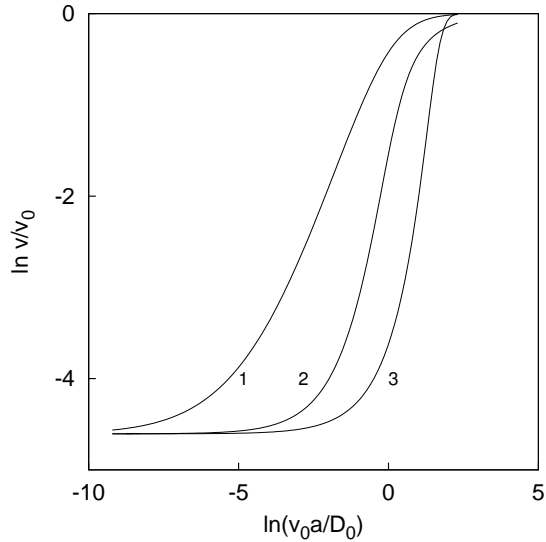}
\caption{Mobility field dependence for various kinds of the correlation function $c^2(x)$: $a^2/(x^2+a^2)$ for curve 1, $\exp(-x/a)$ for curve 2, and $\theta(a-x)$ for curve 3, correspondingly; for all curves $1-\varkappa^2=0.01$.} \label{XY_fig3}
\end{figure}

At last, we consider the exponential correlation function $c^2(x)=\exp(-ax)$ as an example of the function describing correlations with well defined finite length $1/a$ yet still being nonzero for all $x$. For such function the integral in \eq{stat-v_inf} becomes
\begin{equation}
\int\limits_0^\infty dx\frac{\exp(-\gamma x)}{1-\varkappa^2\left[1-\exp(-ax)\right]}=\frac{1}{a}\int\limits_0^1 dy\frac{(1-y)^{\gamma/a-1}}{1-\varkappa^2 y}=
\frac{\Gamma\left(\frac{\gamma}{a}\right)}{a\Gamma\left(\frac{\gamma}{a}+1\right)}
F\left(1,1;\frac{\gamma}{a}+1;\varkappa^2\right),
\label{v-exp}
\end{equation}
here $F(a,b;c;x)$ is a hypergeometric function (Ref. \onlinecite{Gradshtein:book}, integral 3.197.3). For the average velocity $v$ the asymptotics for $v_0\rightarrow 0$ is the same as \eq{2a-v1_E->0}, while
for $v_0\rightarrow \infty$
\begin{equation}
v\simeq v_0-\varkappa^2 D_0 a.
\label{v-exp2}
\end{equation}
Fig. \ref{XY_fig3} shows the behavior of the mobility field dependence for various types of the correlation functions and clearly illustrates a general tendency that the faster is the decay of the spatial correlations of $U$, the steeper is the rise of the mobility field dependence.

\section{Mobility field dependence: critical point $U_0/kT=1$}
\label{critical}

Calculation of the mobility field dependence exactly at the critical point $\varkappa=1$ deserves a special attention. At the critical point the average carrier velocity is
\begin{equation}
v=D_0\left[\int\limits_0^\infty \frac{dx}{c^2(x)}\exp(-\gamma x)\right]^{-1}.
\label{vk1}
\end{equation}
This equation gives us a possibility to draw a clear demarcation line between the short and long range correlation: if $c^2(x)$ decays more slowly than the exponential function (the long range correlation), then the integral (\ref{vk1}) converges and at the critical point $\varkappa=1$ carrier does have a nonzero velocity in the infinite medium. In the opposite case (when $c^2(x)$ decays faster that the exponential function, i.e. in the case of short range correlation) the integral diverges and charge transport is a dispersive one.

For the power law correlation function $c^2(x)=a^n/(x^2+a^2)^{n/2}$ the field dependence of the carrier velocity follows the same dependence $v\propto v_0 (a\gamma)^n$ as in \eq{2a-Sp3}, but for $\varkappa=1$ this dependence is valid for arbitrary small $\gamma\rightarrow 0$ and the Ohmic regime with $v\propto v_0$ does not exists. This means that the so called "nondispersive" regime for the power law correlation function at $\varkappa=1$ significantly differs from the usual nondispersive transport in amorphous materials where for $v_0\rightarrow 0$ the mobility goes to some nonzero constant.

A special case is an exponential correlation function  $c^2(x)=\exp(-ax)$, where a critical bare velocity $v_c=a D_0$ exists. For  $v_0 > v_c$ a nonzero carrier velocity $v=v_0-v_c$ does exists (\eq{v-exp2} becomes an exact relation for $\varkappa=1$), while for $v_0 < v_c$ the carrier velocity goes to 0 for the infinite medium. This phenomenon is in close resemblance to the localization of movable charge carriers in weak field in the the dielectric medium containing equal concentrations of randomly located positive and negative static charges. \cite{Bouchaud:285,Bouchaud:127,Novikov:119}

\section{Mobility field dependence: heuristic approach to the dispersive regime}
\label{dispersive}

Approach, exploited in Section \ref{ND}, cannot be directly implemented in the case of dispersive transport $\varkappa > 1$. A possible direct way should be an explicit averaging of the carrier velocity for a finite thickness $L$ in \eq{stat-v}. We cannot perform this procedure at the moment. Instead, we are going to try a heuristic approach based on the general knowledge of the dynamics of charge relaxation in the exponential DOS. For the long time regime the typical energy of the relaxing carriers (or demarcation energy) is
\begin{equation}
U_d(t)\simeq -3U_0 \ln\left(\frac{B U_0}{kT}\right) -kT\ln\left(\Gamma_0 t\right),\hskip10pt B=\frac{3(4\pi/3)^{1/3} N_0^{1/3}r_0}{2e},
\label{Er}
\end{equation}
where $r_0$ is the localization radius of the transport site, and $\Gamma_0$ is the amplitude of the hopping rate. \cite{Monroe:146,Baranovskii:45} This means that energies much deeper than $U_d(t)$ are not important for the time $t$. Let us introduce a cut-off of the DOS for $U < -U_L\simeq U_d(t)$
\begin{equation}
g_L(U)=N_0\frac{\theta(U_L+U)\theta(-U)}{U_0\left(1-e^{-U_L/U_0}\right)}\exp\left(U/U_0\right).
\label{P_t}
\end{equation}
Using this distribution we can calculate the average time $t_L$ for a carrier to travel across the transport layer with thickness $L$ and then equates this time to the time in \eq{Er}, thus obtaining a self-consistent relation for the calculation of $t_L$.

For the distribution $g_L(U)$ the proper relation between $U$ and the corresponding Gaussian variable $R^2=X^2+Y^2$ is
\begin{equation}
U=U_0\ln\left[1-\left(1-e^{-U_L/U_0}\right)\left(1-e^{-R^2/2}\right)\right].
\label{Ucut}
\end{equation}
Cut-off at $U=-U_L$ removes the divergence in the correlation function $Z(x)$
\begin{equation}
Z(x)=\frac{1}{4\pi^2\left(1-c^2\right)}\int d\vec{R}_1 d\vec{R}_2
\exp\left[-\frac{\vec{R}_1^2+\vec{R}_2^2-2c\vec{R}_1\cdot\vec{R}_2}{2(1-c^2)}\right]
\left[\frac{1-A\left(1-e^{-R_1^2/2}\right)}{1-A\left(1-e^{-R_2^2/2}\right)}\right]^\varkappa,
\label{Zcut}
\end{equation}
here $A=1-\varepsilon$, $\varepsilon=\exp(-U_L/U_0)\propto(\Gamma_0  t)^{-1/\varkappa}\rightarrow 0$ for $t\rightarrow \infty$, and variables $\vec{R}_{1,2}$ are 2D vectors $\vec{R}_i=(X_i,Y_i)$. We consider the case $\varkappa > 1$, where $A\rightarrow 1$, and the denominator in the last multiplier in the integral (\ref{Zcut}) is the most important, because for $A \rightarrow 1$ and $\varkappa > 1$ it provides a divergence for $R_2\rightarrow \infty$. The nominator is not dangerous and we can immediately set $A=1$ here. After integrating over angles we obtain
\begin{equation}
Z(x)=\frac{1}{1-c^2}\int\limits_0^\infty dR_1 dR_2 R_1 R_2
\exp\left[-\frac{R_1^2+R_2^2}{2(1-c^2)}-\varkappa R^2_1/2\right]
\frac{I_0\left(\frac{cR_1R_2}{1-c^2}\right)}{\left[1-A\left(1-e^{-R_2^2/2}\right)\right]^\varkappa},
\label{Zcut2}
\end{equation}
where $I_0(x)$ is a modified Bessel function and after integrating over $R_1$
\begin{eqnarray}
\label{Zcut3}
Z(x)=\frac{K}{1+\varkappa}\int\limits_0^\infty dR_2 R_2
\frac{\exp\left(-KR_2^2/2\right)}{\left[1-A\left(1-e^{-R_2^2/2}\right)\right]^\varkappa}\approx
\frac{K}{1+\varkappa}\int\limits_0^1 dq\frac{q^{K-1}}{(\varepsilon+q)^\varkappa},\\
K=\frac{1+\varkappa}{1+\varkappa\left[1-c^2(x)\right]}.\nonumber
\end{eqnarray}
There is the explicit analytical expression for the integral (\ref{Zcut3})
\begin{equation}
J(\varepsilon)=\int\limits_0^1 dq\frac{q^{K-1}}{(\varepsilon+q)^\varkappa}=
\frac{1}{K\varepsilon^\varkappa}F(\varkappa,K;1+K;-1/\varepsilon)
\label{Zcut4a}
\end{equation}
(Ref. \onlinecite{Gradshtein:book}, integral 3.194.1), but the most useful is the asymptotics for $\varepsilon\rightarrow 0$
\begin{equation}
J(\varepsilon)=\begin{cases}\varepsilon^{K-\varkappa}\frac{\Gamma(K)\Gamma(\varkappa-K)}{\Gamma(\varkappa)},
\hskip10pt \varkappa > K,\\
\ln\left(\frac{1}{\varepsilon}\right),\hskip10pt \varkappa = K,\\
\frac{1}{K-\varkappa},\hskip10pt \varkappa < K.
\end{cases}
\label{Zcut4b}
\end{equation}
For the nondispersive regime $\varkappa < 1$, where $K > \varkappa$, the limit of \eq{Zcut3} for $\varepsilon \rightarrow 0$ gives exactly \eq{Z}.

An average time for a carrier to travel from the reflecting boundary at $x=0$ to the absorbing one at $x=L$ is \cite{Gardiner:book}
\begin{equation}
t_L=\frac{1}{D_0}\int\limits_{0}^{L}dx \int\limits_0^x dx' e^{\gamma(x'-x)}\left<e^{[U(x)-U(x')]/kT}\right>=\frac{1}{D_0}\int_{0}^{L}dx \int_0^x dy e^{-\gamma y}Z(y).
\label{T_avg}
\end{equation}
Here we consider the case of the thick sample  $L\rightarrow \infty$ for the finite $\gamma$, so $\gamma L \rightarrow \infty$ and
\begin{equation}
t_L\simeq \frac{L}{D_0}\int\limits_{0}^{\infty}dy e^{-\gamma y}Z(y),
\label{T_avg_2}
\end{equation}
because the integral over $y$ in \eq{T_avg} converges at
$y\simeq {\textrm{max}}\left(1/\gamma,a\right)$, where $a$ is some characteristic scale for $Z(x)$.

The simplest example is the short range correlation with $c^2(x)=\theta(a-x)$. In this case for $\varkappa > 1$
\begin{equation}
J(\varepsilon)=\theta(a-x)\int\limits_0^1 dq\frac{q^\varkappa}{(\varepsilon+q)^\varkappa}+\theta(x-a)\int\limits_0^1 \frac{dq}{(\varepsilon+q)^\varkappa}\simeq  \frac{\theta(x-a)}{(\varkappa-1)\varepsilon^{\varkappa-1}}
\label{J-shortrange}
\end{equation}
(we keep here the leading term in $1/\varepsilon$) and
\begin{equation}
t_L\simeq \frac{L}{\gamma D_0}e^{-\gamma a}\frac{1}{\varkappa^2-1}\left(B\varkappa\right)^{3(\varkappa-1)}\left(\Gamma_0 t\right)^{1-\frac{1}{\varkappa}},
\label{T-shortrange}
\end{equation}
or, finally, assuming $t_L\simeq t$
\begin{equation}
t_L\simeq\frac{1}{\Gamma_0}\left(\frac{L\Gamma_0 e^{-\gamma a}}{\gamma D_0(\varkappa^2-1)}\right)^\varkappa \left(B\varkappa\right)^{3\varkappa(\varkappa-1)} \label{t-final-short}
\end{equation}
and
\begin{equation}
v_L/v_0\simeq \left(\frac{L\Gamma_0}{v_0}\right)^{1-\varkappa}e^{\varkappa\gamma a}
\frac{(\varkappa^2-1)^\varkappa}{(B\varkappa)^{3\varkappa(\varkappa-1)}}.
\label{v-shortrange}
\end{equation}
Note a typical dependence $\mu\propto (L/v_0)^{1-\frac{U_0}{kT}}$, the same as the corresponding dependence for the dispersive transport in the MT model.  \cite{Rudenko:209} In addition, we get the exponential mobility dependence on the bare carrier velocity $v_0$ (i.e., on the applied electric field $E$). This is the same dependence as in the nondispersive case but the exponent is enhanced by the factor $\varkappa > 1$.

Let us briefly consider a special case $\varkappa=1$. Here the asymptotics of $Z(x)$ is different
\begin{equation}
Z=\theta(a-x)\int\limits_0^1 dq\frac{q}{\varepsilon+q}+\frac{1}{2}\theta(x-a)\int\limits_0^1 \frac{dq}{\varepsilon+q}\simeq  \frac{1}{2}\theta(x-a)\ln \left(\frac{1}{\varepsilon}\right),
\label{Z-kap=1}
\end{equation}
and
\begin{equation}
t_L\simeq \frac{L}{v_0}e^{-\gamma a}\ln \left(\frac{1}{\varepsilon}\right),
\label{tL-kap=1}
\end{equation}
so the self-consistency relation becomes
\begin{equation}
t_L\simeq \frac{L}{v_0}e^{-\gamma a}\ln\left(B^3\Gamma_0 t\right). \label{sc-tL-kap=1}
\end{equation}
Solution for $L\rightarrow\infty$ is
\begin{equation}
t_L\simeq \frac{L}{v_0}e^{-\gamma a}\ln\left(B^3\frac{L\Gamma_0}{v_0}e^{-\gamma a}\right) \label{fin-tL-kap=1}
\end{equation}
or
\begin{equation}
\frac{v_L}{v_0}\simeq  \frac{e^{\gamma a}}{ \ln\left(B^3\frac{L\Gamma_0}{v_0}e^{-\gamma a}\right)}. \label{fin-v-kap=1}
\end{equation}
Mobility dependence on $L$ becomes much weaker in comparison to the dispersive case $\varkappa > 1$. Calculations for other types of $c^2(x)$ become much more difficult, we are going to consider such cases in a separate paper.

\section{Experimental evidence for the correlated exponential landscape}
\label{sec-exp}

Our most reliable results are obtained for the nondispersive transport regime. Obviously, a natural area for the comparison of the theoretical results with the experimental data should be the nondispersive charge transport in amorphous organic semiconductors where the exponential DOS is expected, for example, in amorphous conjugated polymers or oligomers. Unfortunately, the existing experimental data for such regime in the materials in question is very scarce and comparison with the experimental data is a very tedious task because of the necessity to keep the temperature very close to the transition to the dispersive regime. In addition, any mesoscopic inhomogeneities in the experimental sample may lead to the fluctuation of the transition temperature and provide additional complications. Good illustration of the difficulties related to the exact maintenance of the ratio $U_0/kT$ can be found in Ref. \onlinecite{Frost:255}: for one thoroughly studied organic material (P3HT:PCBM blend) the magnitude of $U_0$ varies significantly depending on the morphology of the transport film and experimental method used to estimate $U_0$. Hence, the easiest way to compare our results with the experimental data is to study the mobility field dependence in the dispersive regime thought theoretical results are not so reliable and our analysis is mostly limited to the case of short range correlations.

In comparison to organic materials, much more experimental data is available for the dispersive charge transport in amorphous inorganic semiconductors. For majority of inorganic semiconductors the usual treatment using the MT model (band transport via delocalized states interrupted by frequent trapping to the localized states) probably provides a more adequate description of the transport process, while the hopping transport becomes important for low temperature. At the same time, some authors suggested that the hopping mechanism is more suitable for the description of the dispersive transport in particular amorphous inorganic semiconductors, such as As$_2$Se$_3$, even for rather high $kT$ quite comparable with $U_0$.\cite{Pfister:2062} Computer simulation demonstrates that the waiting-time distribution for carriers in the exponential DOS is practically the same for the MT and hopping mechanism, making the distinction even more problematic.\cite{Hartenstein:8574} Keeping in mind all these complications, we are going to provide a limited comparison of our results with the experimental data.

If we suppose that the short range correlation is indeed the case for amorphous semiconductors with exponential DOS, then at low fields we should expect
\begin{equation}\label{low}
    v_L/v_0\propto \left(\frac{E}{L}\right)^{1/\alpha -1}
\end{equation}
which agrees well with the prediction of the MT model \cite{Rudenko:209} and experimental data, \cite{Pfister:747,Pfister:2062} while for the strong field we should expect an additional exponential factor \cite{Pfister:1147}
\begin{equation}\label{strong}
    v_L/v_0\propto \left(\frac{E}{L}\right)^{1/\alpha -1}\exp\left(\frac{eaE}{\alpha kT}\right).
\end{equation}
The similar exponential factor arises if we take into account the discreet nature of the hopping process, but in that case the correlation length $a$ is replaced by $\rho/2$, where $\rho$ is the distance between neighbor transport sites. Quite probably, the part of the experimentally estimated length factor $\rho_{\rm exp}$ could be related not to the actual distance between transport sites, but to the correlation length, especially in the case of large $\rho_{\rm exp}\simeq 40-50${\AA} (e.g., this is the case of the charge transport in amorphous As$_2$Se$_3$, the material having well established exponential DOS with $U_0\approx 0.05$ eV \cite{Orenstein:23,Pfister:3676,Pfister:1147}). It is worth to note that the most natural explanation for the exponential factor in \eq{strong} is provided by hopping mechanism but not the MT transport.

Experimental data on the temperature dependence of hopping mobility in amorphous semiconductors give us another opportunity to discuss the validity of \eq{v-shortrange}. Typically, in most experimental papers temperature dependence of the dispersive mobility is analyzed in terms of the activation energy $\Delta$, i.e., assuming $\mu\propto \exp(-\Delta/kT)$.
Our relation (\ref{v-shortrange}) does not have exactly that form but if we analyze it in terms of the effective activation energy
\begin{equation}\label{eff-actE}
\Delta_{\rm eff}=-k\frac{\partial \ln\mu}{\partial (1/T)},
\end{equation}
which is suitable in not so wide temperature range, then
\begin{equation}\label{eff-actE-res}
\Delta_{\rm eff}\approx \Delta_0+U_0\left[(6\varkappa-3)\ln\left(B\varkappa\right)+3\varkappa-2\ln\varkappa-5\right],
\end{equation}
here $\Delta_0$ is a microscopic activation energy associated with $\mu_0$ (one should note that $\Delta_0$ totally cancels for the ratio $\Gamma_0/v_0$ in brackets in \eq{v-shortrange}).

We are not going to provide a quantitative comparison of our $\Delta_{\rm eff}$ with experimental $\Delta_{\rm exp}$ for three reasons: first, our treatment of the dispersive regime is very qualitative just to show general trends; second, the estimation of the demarcation energy using \eq{Er} is rather crude, especially the estimation of the numeric coefficient in parameter $B$ (one can compare, for example, estimations of this coefficient in Refs. \onlinecite{Monroe:49,Baranovskii:45}), and, third, we cannot state with assurance that the hopping mechanism (and not the MT mechanism) does provide the dominant contribution in any particular semiconductor. Instead, we will consider the general tendencies which are best demonstrated for the case of highly dispersive transport $\varkappa\gg 1$. \eq{eff-actE-res} demonstrates that the effective activation energy for low field should increase with decreasing temperature. Such behavior has been observed for amorphous As$_2$Se$_3$.\cite{Pai:752}

In conclusion, we have to admit that at the moment it is very difficult to distinguish effect of short range correlation from the effect of the discreet nature of charge hopping. Quite probably, some amorphous organic materials could develop   exponential DOS with long range spatial correlation which should demonstrate more specific transport properties. Indeed, the case of the short range correlated disorder is, probably, the most insensitive to the particular properties of the amorphous transport material; we already noted that, for example, the mobility field dependence for that case is essentially the same for the Gaussian and exponential DOS. We have to note also that the comparison of the experimental and calculated mobility field and temperature dependences could be not so straightforward task for organic materials because in some materials the experimental data suggest that $U_0$ itself depends on $T$. \cite{Street:165207} Additional complication is a possible unfinished energetic relaxation of charge carriers under typical experimental conditions in modern organic electronic devices. Strictly speaking, our results are valid in the limit $L\rightarrow\infty$, while in modern organic electronics there is a clear tendency to use very thin transport layers.

\section{Conclusion}

We considered 1D model of the charge carrier transport in the highly disordered amorphous semiconductors having the spatially correlated exponential localized DOS and predicted the transport behavior for such materials. Amorphous organic semiconductors with rather weak intermolecular interaction and low dimensional disordered materials where the disorder leads to the localization of all relevant states are favorable for the realization of the discussed hopping transport mechanism.
The exact formula for the average carrier velocity in the infinite medium was obtained for the nondispersive quasi-equilibrium transport regime. Consideration of the charge transport exactly at the transition temperature to the dispersive regime provides the explicit criterium for the distinction between the short range and long range correlation. At the same time, our study provides a lot of open questions for further investigation.

The most important question is related to the Gaussian representation used in our study. It was already mentioned that not all correlated exponential distributions can be generated in this way. For example, a distribution having binary correlation function $c_U(x)$ which is negative for some $x$ cannot be generated using our approach. Probably, such correlation functions are less typical for amorphous materials but, nonetheless, they are not unphysical. A simple example of such correlation function, though not in the case of the exponential distribution, could be provided by the random energy landscape, generated by randomly oriented dipoles located at the sites of the 1D line \cite{Novikov:14573} or for the disordered material with site energies given by the charge-induced dipole interaction. \cite{Freire:134901} In addition, we cannot guarantee that even all distributions having non-negative $c_U(x)$ may be constructed using the Gaussian representation. The natural question is: what are physical effects (if any) of the non-Gaussian nature of the spatial correlation for non-negative $c_U(x)$? Does the mobility field dependence differ qualitatively from relation provided by \eq{stat-v_inf} and \eq{Z}? We are going to address this problem in future.

A very actual open problem is what kind of correlation functions  could be observed in real amorphous materials having the exponential DOS. At the moment we have no reliable information about spatial correlations in inorganic amorphous materials. We provided a brief comparison of our results with experimental data for amorphous As$_2$Se$_3$ assuming the simplest short range correlation of random energies. Unfortunately, the discreet nature of hopping transport (the final distance between neighbor transport sites) provides very similar mobility field dependence. Probably, organic materials with the exponential tail of the DOS are more promising for the experimental study of the correlation effects. Indeed, organic materials typically demonstrate long range spatial correlation of the energy landscape. Yet at the moment spatial correlations in organic amorphous materials with exponential DOS are not studied.

Another interesting problem is a more detailed characterization of the various "nondispersive" regimes mentioned earlier. In this paper we used a very simple definition of the nondispersive charge transport as a transport regime where the constant time-independent average carrier velocity eventually develops for the carrier traveling in the infinite medium. For such regime the carrier velocity measured in the time-of-flight experiment for the sufficiently thick transport layer does not depend on thickness.  Yet the very term "dispersive" was coined in relation to the shape of the current transient being monotonously decaying and featureless if plotted in double linear coordinates current vs time. Common use of the term "dispersive" is generally related to transients decaying as $\propto t^{-1-\alpha}$ for $t\rightarrow\infty$ ($0< \alpha < 1$). Results, presented in this paper, do not shed any light on the shapes of the transients. We may expect that shapes of the transients for the regimes described in Sections \ref{critical} and \ref{dispersive} are not identical. Direct calculation of the shape of the transient for the hopping charge transport is a very difficult task, but we may expect that a very important information about the shape may be obtained by the calculation of the diffusivity. Validity of the Einstein (or modified Einstein \cite{Parris:2803}) relation should be verified as well. This is an additional task for the future research.

Our treatment of the dispersive transport regime is very approximate, at best. More accurate consideration requires the calculation of the carrier velocity for the transport layer with finite thickness, and this is a much more formidable task. This problem will be considered in future. Finally, we would like to note that our approach is not limited exclusively to the case of the exponential DOS. Mapping to the Gaussian distributions according to various approaches discussed in Ref. \onlinecite{Devroye-book} may be extended to other shapes of the DOS.

We may conclude that there is a lot of interesting and important open physical problems associated with the hopping charge transport in the spatially correlated exponentially distributed random energy landscape.

\section*{Acknowledgements}
Financial support from the Russian Science Foundation grant 15-13-00170 is acknowledged.

\end{document}